\documentstyle[12pt,moriond,psfig]{article}
\begin{document}

%
%

\heading{ELLIPTICAL GALAXY DYNAMICS: THE ISSUES PERTAINING TO GALAXY FORMATION}

\author{David Burstein $^{1}$}  {$^{1}$ Department of Physics and 
Astronomy, Arizona State University, Tempe, AZ USA 85287-1504.} {} {} {}

\begin{moriondabstract}
How elliptical galaxy dynamics relate to galaxy structure, stellar populations,
spiral galaxies and environment are reviewed.  The evidence assembled shows
that most, if not all, galaxies originally classified as gE contain disks
within them.  Taken together, the existing evidence are most consistent with
the gravitational, hierarchical, clustering, merging (HCM) concept that all
galaxies, including gE, are formed from the combination of much smaller
galaxies.  Within the HCM picture, the evidence also strongly suggests that
the subunits which go into forming galaxies must be related in specific ways
both to the mass of the galaxy they will form, and to the environment in which
the galaxy forms.  Despite the extensive data that we now have on galaxies
that can constrain their formation history, the lack of a physical
understanding of the stellar initial mass function prevents us from developing
realistic physical models for galaxy formation.

\end{moriondabstract}

\section{Introduction}

Recent reviews on the dynamics of elliptical galaxies {\it per se} have been
given in the last several years by \cite{Me99}, \cite{deZ96}, \cite{St95},
\cite{Ba96}, \cite{Ge94}. Each of those reviews covers a difference aspect of
the wealth of information we now have, both observational and theoretical, on
the detailed structures of elliptical galaxies.  Yet, that informaton is only
part of the story we need to know if we are ever to fully understand galaxy
formation in general, and the formation of elliptical galaxies in particular.
To this detailed information we have to add evidence on how the dynamics of
ellipticals are related to their stellar populations, to the dynamics of
spiral galaxies, and to their environment.  To this evidence we have to also
understand the physics that underly star formation in general, and the
existence of a well-defined stellar initial mass function in particular. As we
will discuss in this review, even what we call an ``elliptical galaxy'' is no
longer clearly defined, as detailed photometric and dynamical studies are
finding disks in most ``giant elliptical'' galaxies.

This review uses existing evidence in the literature, supplemented by the
results from the author's participation in the EFAR survey (\cite{Weetal96},
\cite{Weetal99}, \cite{Saetal96}, \cite{Saetal97}, \cite{Coetal99}), and from
the existing information on the distribution of nearby galaxies. The
discussion will be divided into six parts:  a brief summary of the dynamics
and structure of giant elliptical (gE) and S0 galaxies; the relationship of
stellar populations to dynamics and structures of E galaxies in general;
presence or absence of disks in gE galaxies plus some of the kinds of disks
seen in gE galaxies; the relationships among gE, S0, spiral galaxy dynamics;
the environments in which we find gE and S0 galaxies near us; and a discussion
of what new perspectives we might extract from this evidence regarding galaxy 
formation.

\section{Dynamics and Structure}

Ever since it was discovered that giant elliptical (gE) galaxies do not rotate
to support their observe flatness (e.g., \cite{BeCa75}), their internal
dynamics have been interpreted in terms of anisoptropic velocity dispersions
resulting in intrinsically triaxial figures (e.g., \cite{Bi78}). As Merritt
(\cite{Me99}) summarizes in the most recent extensive review on this subject,
our knowledge of the dynamical structure of gE galaxies is based both
on theoretical models of this structure and on detailed studies of individual
galaxies.

All of these studies indicate that gE galaxies can have a wide variety of
intrinsic structure, ranging from mostly prolate to mostly oblate, but that
the characterization of their shapes as triaxial is most general. The
interested reader is referred to the fine reviews on this subject for further
details.

\section{Dynamics and Stellar Population}

That the global optical parameters of gE galaxies --- $\rm r_e$, $\rm I_e$ 
--- form a plane within the volume defined in conjunction with central
velocity dispersion ($\rm \sigma_c$) is {\it prima facie} evidence that a
strong relationship must exist between the stellar populations of gE galaxies
and their dynamics (cf. discussion by Renzini \& Ciotti \cite{RC93}).

Bender, Burstein \& Faber (\cite{B2F93}) extended this evidence to dwarf
elliptical and dwarf spheroidal galaxies by comparing central measures of the
line index $\rm Mg_2$ and $\rm \sigma_c$ for gE, dE and spiral bulges with
more global values of [Fe/H] (transforming this to $\rm Mg_2$ via the
relationship given in \cite{Buetal84} for Galactic globular clusters) and $\rm
\sigma$ for dSph. As shown in Figure 3 of \cite{B2F93}, the relationship $\rm
Mg_2 = 0.20 \log \sigma - 0.l66$ fits well all of these early-type systems,
from the $10^{12}$ M$_\odot$ gE galaxy to the $10^{5}$ M$_\odot$ dSph. 

Colless et al. {\cite{Coetal99} used our EFAR data to show that the $\rm
Mg_2$--$\sigma$ relationship is the same, within the errors, among 423
galaxies in 84 different galaxy groups and clusters. Moreover, in that paper
it is possible to derive an {\it intrinsic} variation of $\rm Mg_2$ on
$\sigma$ (as opposed to error-caused) of 0.016 mag.

Without going into the subtleties of how the $\rm Mg_2$ index is related to
stellar population, these empirical relationships establish a strong
connection among the stellar populations of galaxies spanning a range of $\rm
10^7$ in mass. To this writer, the fact that all of these ``dynamically hot''
stellar systems have stellar populations related to their internal dynamics
{\it and to one another} is one of the central puzzles that any theory of
galaxy formation must solve to be succcessful. There are currently two ideas
that have been proposed as an overall framework to interpret these correlations
(modulo our current lack of understanding of the physics of forming the
stellar initial mass function):

Franx \& Illingworth (\cite{FrIl90}) suggest that the central gravitational
potential determines the stellar population of a galaxy. However, this would
require that stellar populations of gE galaxies be dependent on both $\sigma$
and $\rm r_e$, a prediction not consistent with the study of \cite{B2F93}.
From their analysis, Bender et al. (\cite{B2F93}) derive that the stellar
population is related to both the mass density ($\rho$) of the galaxy {\it
and} to its overall mass (M), in the form $\rm M^2 \rho$. Bender et al. use
this relation to predict that stellar population gradients within galaxies are
small ($\rm \delta(B-V) \propto 0.037 \log \rho$). Unfortunately, such a
prediction is hard to confirm, as small stellar population gradients are
difficult to measure in the face of systematic problems in the measurements
(e.g., differing seeing radii on images in different passbands; sky
subtraction issues).

\psfig{file=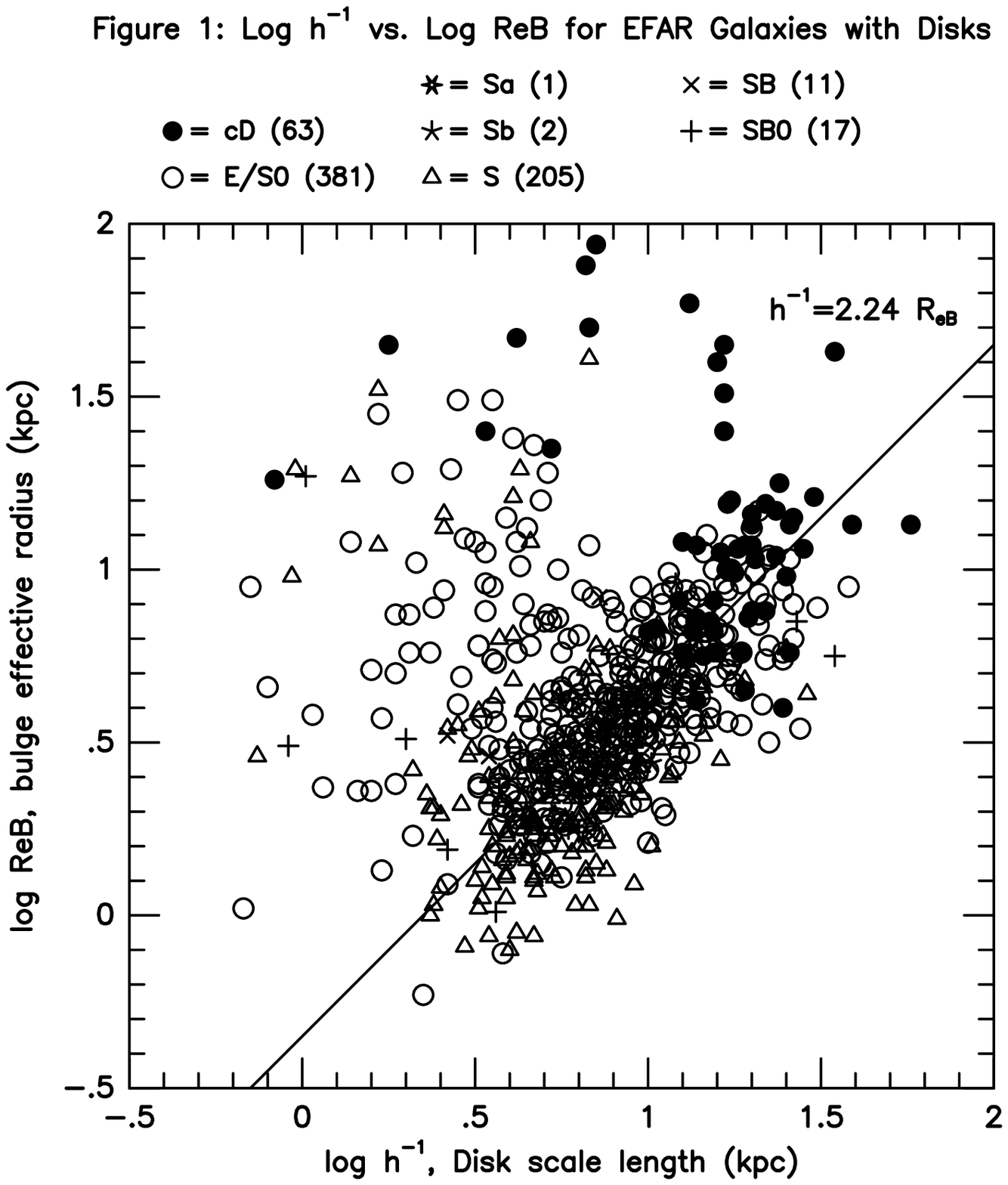}

Also related to this issue is the fact that \cite{Scetal90} find the strength
of the Balmer absorption line $\rm H\beta$ in the spectra of gE galaxies is
related to optical evidence of recent interactions with other galaxies.  This
leaves open the issue that any stellar population disperion among gE galaxies
at a given dynamical parameter can be due to residual effects of the most
recent star formation event linked to the most recent interaction event.

\section{The Paradigm Shifts: Find a gE Galaxy without a Disk!}

The separation of galaxies into gE and S0 classes is based primarily on the
optical appearances of the galaxies.  By these morphological definitions, S0
galaxies are diffuse in appearance (i.e., no obvious signs of active star
formation) and show evidence of a disk; gE galaxies show no optical evidence
of a disk, but are otherwise similar in appareance as S0's.

As pointed out by the first systematic study of S0 disk--to--bulge ratios
(\cite{Bu78}, \cite{Bu79}), one can hide a disk within a gE galaxy if the
disk--to--bulge ratio is small enough (typically, 0.1 or less) and the disk is
not oriented edge-on to the line--of--sight.  However, it was not until the
advent of CCD imaging that the signal-to-noise of galaxy images became high
enough that faint disks in morphologically-classified gE galaxies could be
easily found from direct images.

Davies et al. (\cite{DEFIS83}) gave the first hint that many gE galaxies
contain disks by pointing out that the ratio of rotational-to-anisotropic
motions in gE galaxies is related to absolute luminosity:  The proportion of
galaxies that are primarily supported by anisotropic velocity dispersions
increases with increasing galaxy luminosity. R. Bender and his colleagues
(\cite{BeMu87}, \cite{BDM88}) then used the more accurate CCD data
to define shape parameters for a large sample of gE galaxies via deviations of
their image from a pure elliptical shape. Based on these kinds of analyses,
both they and others have shown we can divide gE galaxies into two general
classes:  boxy and disky.  As shown in a number of studies (cf. \cite{B2F92},
\cite{B2F93}), disky gE galaxies tend to be less luminous than boxy gE
galaxies, albeit with a substantial amount of crossover.

Saglia, Bender \& Dressler {\cite{Sa93}) found that gE galaxies with
isophotes classifed as ``disky'' define fundamental planes parallel to, but
offset towards slightly lower M/L from that defined by ``boxy'' isophotes.
They interpret this offset as due to the importance of rotation in determining
the mass of disky galaxies which is not taken into account by just using central
velocity dispersion for the mass calculation.

Our EFAR survey {\cite{Saetal96}, \cite{Saetal97}) obtained CCD and
photoelectric photometry for 776 galaxies originally classified by eye on
the Palomar Sky Survey prints as being gE candiates with axial ratios mostly
face on (b/a $>$ 0.4).  Of the 537 non-barred early-type galaxies in this
sample (the sample was purposefully oversampled with spiral galaxies), 444
(83\%) are classified as photometrically having evidence of a disk: E/S0, S0
or cD. In almost all cases this was done with mulitple images for each galaxy,
via a careful disk/bugle deconvolution procedure that properly took into
account seeing radius and sky subtraction issues (cf. \cite{Saetal96},
\cite{Saetal97}). The reader is referred especially to the 32 graphs of
luminosity profiles for each galaxy in our survey presented in \cite{Saetal97}.

The extensive simulations done in \cite{Saetal96} show that for any gE galaxy
in which the disk is of similar surface brightness as the bulge, but of less
than 10\% the luminosity of the bulge, the disk is essentially invisible in
the data we have.  This is similar to the conclusion reached by \cite{Bu78}
from analyzing the data on S0 galaxies.

To go one step further, in our EFAR data we have looked for correlations among
the disk and bulge parameters we have derived for these galaxies. The most
interesting correlation we have found is between the exponential scale length
of the disk ($\rm h^{-1}$ from the definition $\rm I(r) \propto e^{\rm -hr}$)
and the de Vaucouleurs R$^{1/4}$-law effective radius of the bulge ($\rm
R_{\rm eB}$) for those galaxies in our sample that have disks: 63 cD's, 381
S0's, 17 SB0's, 208 non-barred spirals, and 11 barred spirals - 680 disk
galaxies in all.  Figure~1 shows this relationship for these galaxies, with
different plotting symbols used for each galaxy type, as indicated in the
figure.

As is evident, most of the 652 unbarred, disky galaxies, including all galaxy
types (cD, E/S0, S0 and spiral), show a good relationship between disk scale
length and bulge effective radius of the form $\rm h^{-1} \propto 2.24 R_{\rm
eB}$ (with effective disk radius, $\rm R_{\rm eD} = 1.68 h^{-1}$ for an
exponential disk).  This is consistent with what has been found from other
studies of bulge and disk parameters for late-type spiral galaxies 
(\cite{Coetal96}).

The larger scatter about this relationship is preferentially to one side for
all galaxies --- smaller disk scale lengths for a given bulge effective
radius.  Based on the scatter in the general relationship to the opposite side
(i.e., large disk size for a given bulge size), we separate out 73 disks in
non-barred, non-cD galaxies that are more that 0.45 dex too small for their
bulge sizes from the rest of the sample. We will tentatively term these disks
as ``interior disks,'' as we find them in the interiors of galaxy bulges.
These 73 galaxies are close to 10\% of the 745 non-barrred galaxies in our
whole sample (i.e., including 88 gE and 5 cD galaxies in which we see no
photometric evidence of a disk).

It is also evident upon inspection that these ``interior disks'' have
preferentially small disk scale lengths when present, generally 3 kpc or less.
Further investigation shows that ``interior disks'' have systematically higher
surface brightnesses, and reside in galaxies with systematically fainter bulge
surface brightnesses, than do disks and bulges of galaxies whose bulge and
disk scale lengths fit the general relation.  The combination of high surface
brightness interior disk and low surface brightness bulge makes the interior
disks easier to find. Thus, our photometry has uncovered evidence that at
least 10\% of E, E/S0, S0 and spiral galaxies contain within them small, high
surface brightness disks, including five galaxies (NGC~4887, NGC~4839,
Dressler~136, IC~4051 and IC~4052) in the Coma cluster.  Significantly, one of
these Coma cluster galaxies (IC~4051) has been shown by \cite{Meetal98} to
have a kinematically distinct, high surface brightness disk well inside its 
bulge.

\psfig{file=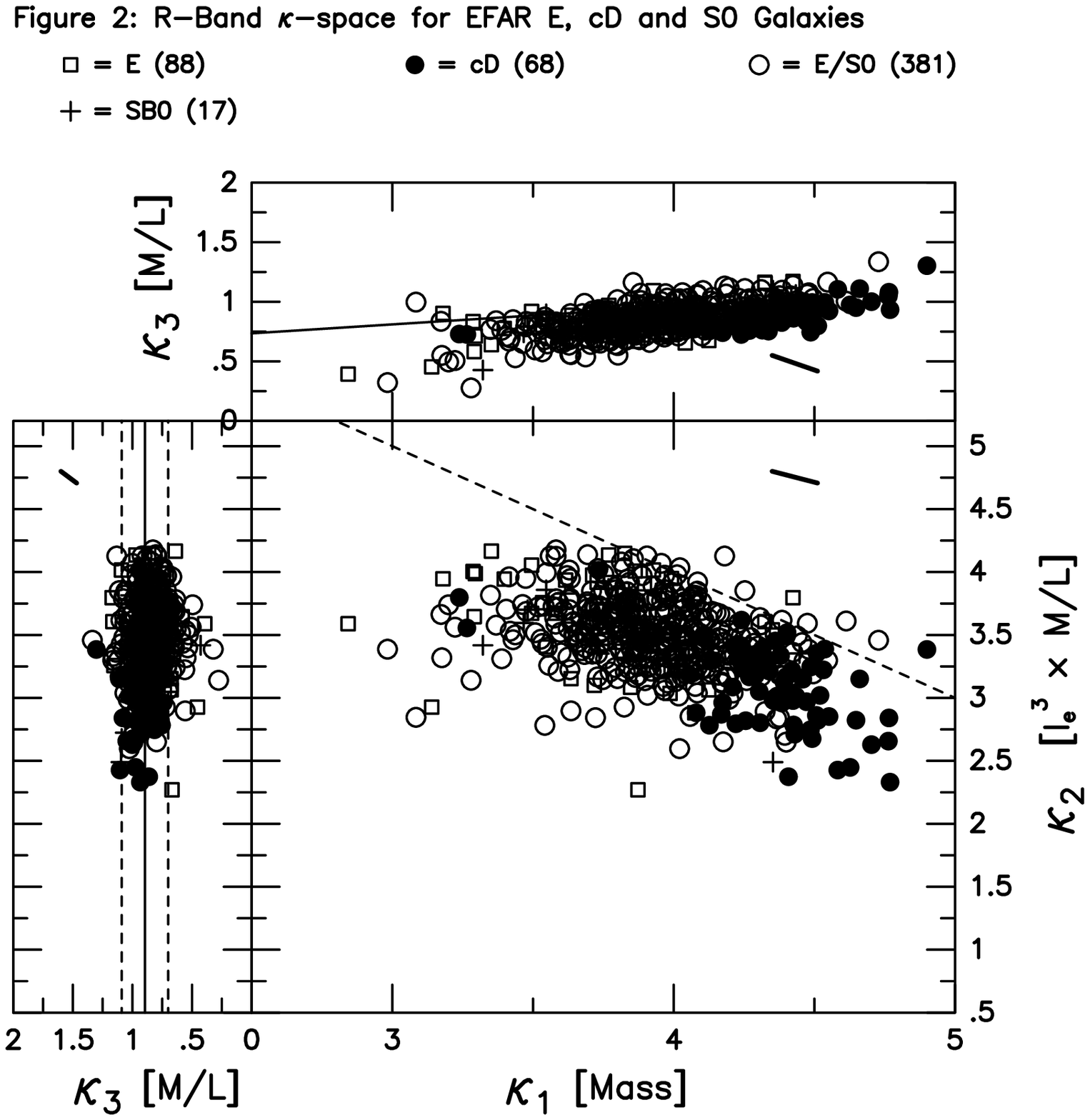}

The EFAR survey preferentially selected more face-on galaxies as gE
candidates, yet found nearly 85\% of them have disks that can be detected
photometrically.  Given that the remaining gE galaxies {\it without} detected
disks could still have disks contributing less than 10\% of their
luminosities, it is highly possible that {\it all gE galaxies contain disks.} 
If this is so, then as the title of this section states, the paradigm shifts
from ``find a gE galaxy with a disk'' to ``prove that any gE galaxy {\it
does not have} a disk!''

\section{Elliptical Galaxy and Disky Galaxy Dynamics are Likely Related}

One can unambiguously define the same set of three global parameters for
all galaxies that go into defining the elliptical galaxy fundamental plane:
effective radius, effective surface brightness and central velocity disperison.
As shown in \cite{B2FN97}, when you define these three global properties for 
spiral galaxies, irregular galaxies and elliptical galaxies in a self-consistent
manner, we find that the physical properties of these galaxies are related
to each other.  

\psfig{file=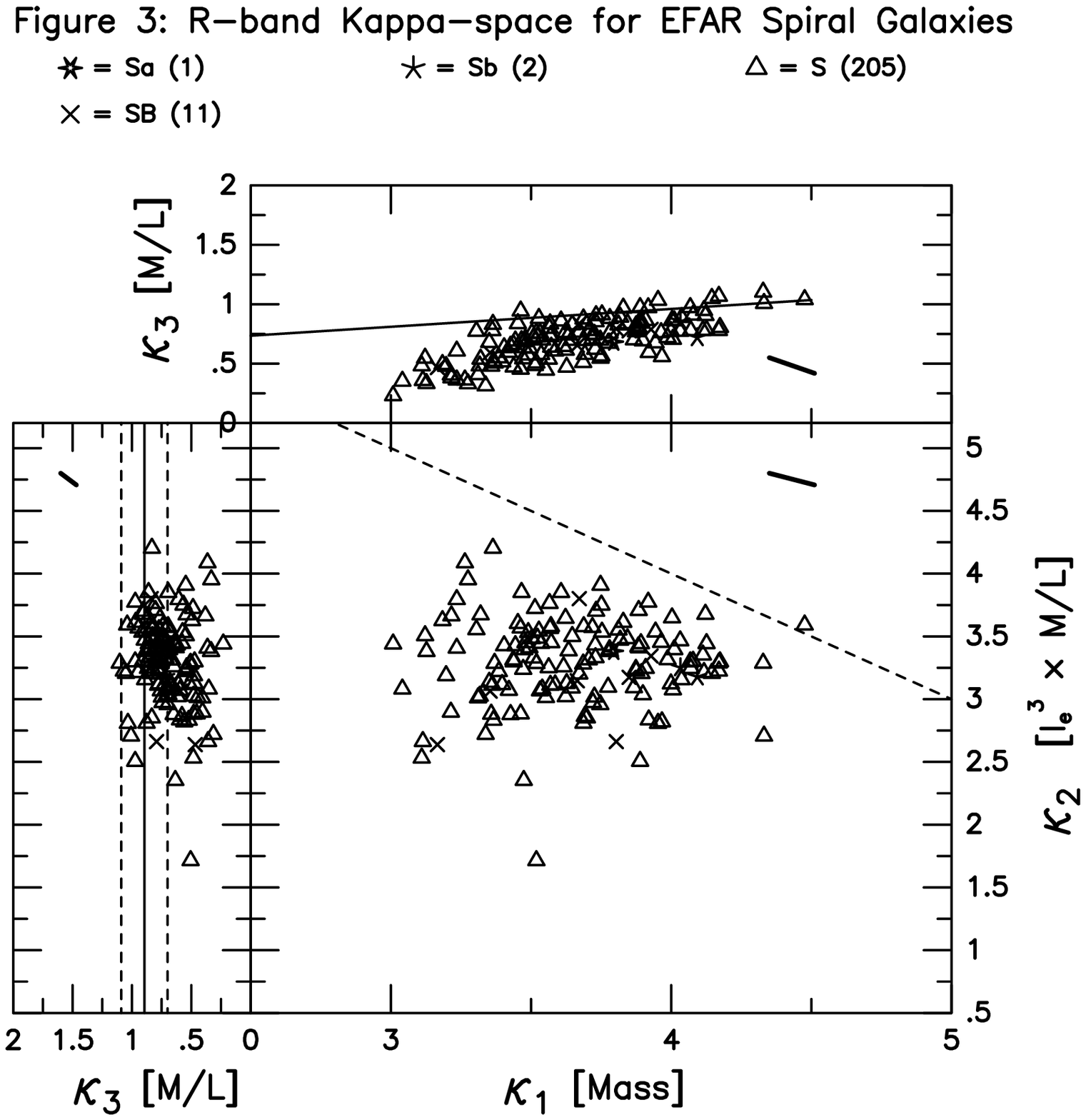}

Burstein et al. (\cite{B2FN97}) used the exising B-band data for these kinds
of galaxies to define their relationships within $\kappa$--space. In Figures~2
and 3 we graphically show the analogous relationships we form from the EFAR
R-band data for the gE, E/S0, cD and spiral galaxies in our sample. (The
number of galaxies per type is different from that given in Figure 1 owing to
the additional requirement of velocity dispersion measurements.) As first
discovered by \cite{B2FN97}, galaxies of different morphological class define
distinct sequences with $\kappa$--space for B-band luminosity-related
properties.  Figures~2 and 3 show that these sequences are also seen within
R-band-defined $\kappa$--space.  The small solid lines drawn in each plane
represent the effect of $\pm$30\% distance errors.  The solid line draw in the
$\kappa_3$--$\kappa_1$ plane is the edge-on view of the fundamental plane in
the B-band.  The dashed line in the $\kappa_2$--$\kappa_1$ is the leading edge
of the zone of exclusion seen in the B-band (cf. \cite{B2FN97}). As with the
B-band data, the fundamental plane is seen most edge-on in the R-band
$\kappa_3$--$\kappa_1$ (M/L vs. M) plane and a well-defined zone of exclusion
of the form $\kappa_1 + \kappa_2 =$ const exists within the R-band
$\kappa_2$--$\kappa_1$ plane.  Comparing Figures 2 and 3, we also see that, as
in the B-band data, spiral galaxies in the $\kappa_2$--$\kappa_1$ plane lie
parallel to, but farther away from the zone of exclusion line than do
earlier-type galaxies.  As discused in detail in \cite{B2FN97}, this evidence
strongly suggests that all galaxies, gE to irregular, are formed in similar
ways.

\section{Nearby Galaxies are Morphologically-Segregated on 15-Mpc Size Scales}

It has been well known for many years that gE galaxies tend to be found within
highly populated, dense clusters.  This trend was first quantified by 
\cite{Dr80}, who showed that the percentage of early--to--late type galaxies
is a function of the local density of galaxies around them.  Early-type galaxies
tend to dominate in number in high density regions of space, while 
late-type galaxies tend to dominate in number in low density regions.

Using data from the {\it Nearby Galaxies Catalog} of Tully ({\cite{Tu88}), we
have selected three regions of size 1000--1500 km/sec (14--20 Mpc if $\rm H_0
= 70$ km sec$^{-1}$ Mpc$^{-1}$) near our Local Group --- the Coma-Sculptor
cloud (in which our Local Group resides), the main part of the Virgo cluster,
and the Ursa Major Cluster.  Using the Hubble types given in the Tully catalog,
we calculate the fraction of {\it giant} galaxies (i.e., not dwarfs) as found
in each of these three regions of space, divided by type into five classes:
E--S0/a, Sa--Sb, Sbc--Sd, Sd--Irr and Peculiar. The results are shown in terms
of percentages of each type in Figure~4.

It is evident that late-type galaxies dominate the whole of the Coma-Sculptor
Cloud, which is about 1500 km sec$^{-1}$ long and very narrow. In contrast,
early-type galaxies comprise more than half of the giant galaxies in the Virgo
cluster, while Ursa Major galaxies distribute themselves rather evenly among
all five morphological classes. In all three cases, the morphological types of
galaxies are related to each other over wide regions of space.

\psfig{file=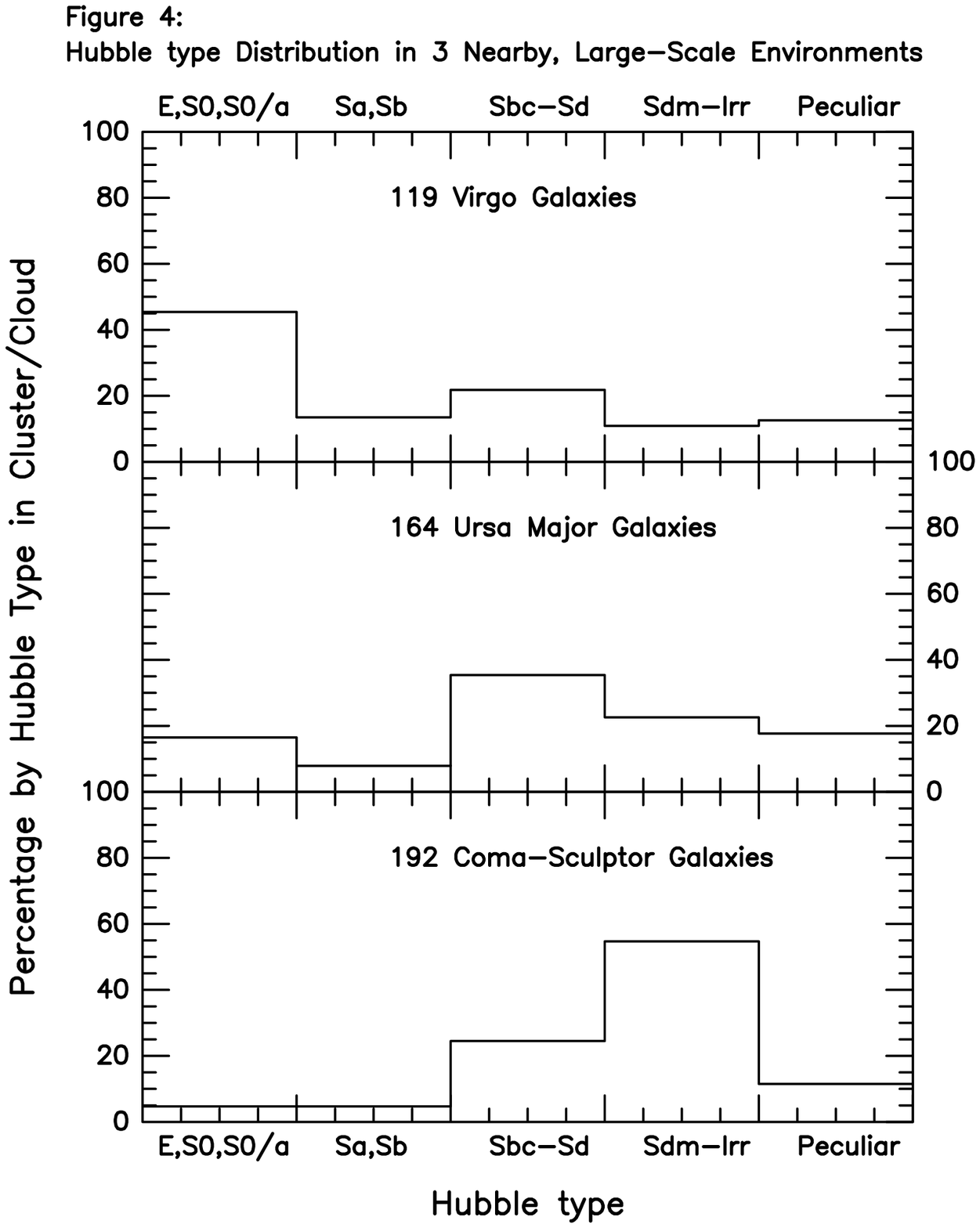}

\section{Discussion}

The current paradigm for forming galaxies is that of hierarchical, clustering,
merging (HCM) growth of galaxies with increasing age of the Universe. In this
picture, the initial density fluctations in the early universe grow
gravitationally from small masses to large masses as the Universe expands. As
discussed in \cite{B2FN97}, the distribution of the global properties of
galaxies within $\kappa$--space is consistent with an HCM interpretation, if
one associates the initial density of a galaxy with a specific range of Hubble
types.  In this picture, denser initial densities lead to earlier Hubble
types, on average. The HCM scenario is also consistent with what we know about
the internal structures of E and S0 galaxies, if one allows (cf. \cite{B2F92})
the more luminous gE galaxies to form from mostly stellar mergers, while less
luminous gE galaxies form from more gaseious mergers.  Thus, the HCM scenario
naturally leads to the observed relationships among galaxy dynamics from gE
to Irr.

However, the HCM paradigm tells us nothing of how to change gas into stars, so
cannot tell us the origin of the tilt of the galaxy fundamental plane in M/L
vs. L, nor why stellar population should be so tightly related to galaxy
dynamics in early-type galaxies. It also tells us nothing about any {\it a
priori} initial conditions in the Universe that might lead to large-scale
(1000-1500 km sec$^{-1}$ sizes as seen today) density fluctuations.  In the
face of current lack of input physics into both of these issues, we can 
speculate on how these relationships might be made.

This author believes that the stellar population--dynamical relation among
early-type galaxies over a range of $10^7$ in galaxy mass pins the initial
subunit size in the HCM picture to be, at most, 1/2 to 1/5 the size of the
smallest dSph, or about 10$^5$ M$_\odot$ in size.  One possible scenario to
generate the observed stellar population-dynamics relationship (certainly
not the only one!) is this: A typical dSph will have 5-10 of these subnits in
it, while a typical gE has 10$^5$--10$^6$ subunits in it.  If we then link the
density ranges of the subunits to the mass of the galaxy in which they will
reside (more mass, higher mean subunit density), and the stellar populations
of the subunits to their densities, we can get the observed relationships.  If
this scenario is correct, it would also require that the subunits do not
arbitrarily combine to form galaxies, but that most subunits ``know'' the kind
of galaxy that they will combine to form.

The close spatial proximity and large-scale differences in galaxy types and,
by the HCM paradigm, in initial mass densities, is a challenge for N-body
modellers to reproduce.  If these features in the observable universe can
be reproduced in existing models, it will be triumph for those models.  If,
however, the models cannot reproduce such large-scale variations in mass
density, one is left looking for other answers.

Finally, lest one be left with the impression that we have covered all issues
relevent to galaxy formation here, consider just two other pieces of
information we have to add to the mix: ubiquitous central supermassive
blackholes in bulges (e.g., \cite{Maetal98}, \cite{Rietal98}), and
thermally-emitting X-ray gas of mass generally comparable to that in the
galaxies themselves existing within galaxy clusters and groups (e.g.,
\cite{Muetal96}).

Taken together, our knowledge of galaxy properties and their relationships to
each other and to their environment has now become so detailed that we can
well constrain any proposed model of galaxy formation.  However, we lack a key
ingredient as important to understanding galaxy formation as nuclear fusion
was to understanding stellar evolution:  Why a well-defined stellar initial
mass function should exist, and the physics that is required to produce it.
In absence of the knowledge of such physics, our modeling of galaxy formation
will continue to be incomplete.

\acknowledgements

This author would like to thank the organizers and hosts of this meeting for
their kind generosity and warm hospitality.  These were much appreciated. The
work reported on here, and in which this author participated, was highly
collaborative in nature, supported in most part by several U.S. NASA and NSF
grants over a long time period.  This author would particularly like to thank
his collaborators in the $\rm B^2F$ papers --- Ralf Bender, Sandy Faber,
Richard Nolthenius; and in the EFAR program --- Roberto Saglia, Matthew
Colless, Roger Davies, Bob McMahan, Glenn Baggley,  Gary Wegner, Ed
Bertschinger, Roelof deJong, for their great help in producing the scientific
results presented here.

\vfill
\end{document}